\documentclass[final]{cvpr}

\usepackage{times}
\usepackage{epsfig}
\usepackage{graphicx}
\usepackage{amsmath}
\usepackage{amssymb}

\usepackage{subfigure}
\usepackage{comment}
\usepackage[misc]{ifsym}

\usepackage[pagebackref=true,breaklinks=true,colorlinks,bookmarks=false]{hyperref}

\def\cvprPaperID{3985} % *** Enter the CVPR Paper ID here
\def\confYear{CVPR 2021}

\begin{document}

%%%%%%%%% TITLE
\title{FVC: A New Framework towards Deep Video Compression in Feature Space}

\author{Zhihao Hu\\
Beihang University\\
{\tt\small huzhihao@buaa.edu.cn}
\and
Guo Lu$^{\ast}$\\
Beijing Institute of Technology\\
{\tt\small guo.lu@bit.edu.cn}
\and
Dong Xu\\
University of Sydney\\
{\tt\small dong.xu@sydney.edu.au}
}

\maketitle

\pagestyle{empty}
\thispagestyle{empty}

\begin{abstract}
{\let\thefootnote\relax\footnote{{$^{*}$~Guo Lu is the corresponding author.}}}
Learning based video compression attracts increasing attention in the past few years. The previous hybrid coding approaches rely on pixel space operations to reduce spatial and temporal redundancy, which may suffer from inaccurate motion estimation or less effective motion compensation. In this work, we propose a feature-space video coding network (FVC) by performing all major operations (\textit{i.e.}, motion estimation, motion compression, motion compensation and residual compression) in the feature space. Specifically, in the proposed deformable compensation module, we first apply motion estimation in the feature space to produce motion information (\textit{i.e.}, the offset maps), which will be compressed by using the auto-encoder style network. Then we perform motion compensation by using deformable convolution and generate the predicted feature. After that, we compress the residual feature between the feature from the current frame and the predicted feature from our deformable compensation module. For better frame reconstruction, the reference features from multiple previous reconstructed frames are also fused by using the non-local attention mechanism in the multi-frame feature fusion module. Comprehensive experimental results demonstrate that the proposed framework achieves the state-of-the-art performance on four benchmark datasets including HEVC, UVG, VTL and MCL-JCV. 
\end{abstract}

\section{Introduction}

% \begin{figure}[!t]
% \centering
% \begin{minipage}[t]{0.5\linewidth}
% \centering
% \subfigure[Original Frame]{\includegraphics[width=0.95\linewidth]{figures/intro/input.png}\label{fig:introframe}}
% \end{minipage}%
% \begin{minipage}[t]{0.5\linewidth}
% \centering
% \subfigure[Reconstructed Optical Flow]{\includegraphics[width=0.95\linewidth]{figures/intro/flow.png}\label{fig:introflow}}
% \end{minipage}
% \begin{minipage}[t]{\linewidth}
% \centering
% \subfigure[]{\includegraphics[width=0.95\linewidth]{figures/intro/compare.pdf}\label{fig:compare}}
% \end{minipage}
% \caption{Failure cases of motion compensation from optical flow based method DVC and the corresponding results of our proposed method FVC (Ours).}
% \label{fig:intro}
% \end{figure}

There is an increasing research interest in developing the next generation video compression technologies. While the traditional video compression methods~\cite{wiegand2003overview, sullivan2012overview} have achieved promising performance by using the hand-designed modules (\textit{e.g.}, block-based motion estimation, Discrete Cosine Transform (DCT)) to reduce spatial and temporal redundancy, these modules cannot be optimized in an end-to-end fashion based on large-scale video data.
% Due to the powerful representation ability of neural networks, deep video compression has become a promising new research direction.

The recent deep video compression works~\cite{lu2019dvc,wu2018video,Agustsson2020space,abdelaziz2019neural,hu2020imporving,lu2020content,liu2020mlvc} have achieved impressive results by applying the deep neural networks within the hybrid video compression framework. 
% For example, Lu \textit{et al.} proposed the DVC~\cite{lu2019dvc} framework, where a pixel-level optical flow estimation network~\cite{ranjan2017optical} is used for motion estimation, while two image compression networks~\cite{balle2016end, balle2018variational} are used for pixel-level optical flow compression and residual compression, respectively. 
Currently, most works only rely on pixel-level operations~(\textit{e.g.,} motion estimation or motion compensation) for reducing redundancy.
For example, pixel-level optical flow estimation and motion compensation are used in \cite{lu2019dvc,lu2020content,hu2020imporving,lu2020anendtoend} to reduce temporal redundancy, and pixel-level residual is further compressed by using the auto-encoder style network. 
However, this pixel-level paradigm suffers from the following three limitations. 
First, it is difficult to produce accurate pixel-level optical flow information, especially for videos with complex non-rigid motion patterns. 
Second, even we can extract sufficiently accurate motion information, the pixel-level motion compensation process may introduce additional artifacts. 
Third, it is also a non-trivial task to compress pixel-level residual information.

Given robust representation ability of deep features for various applications, it is desirable to perform motion compensation or residual compression in the feature space and more effectively reduce spatial or temporal redundancy in videos. 
% Furthermore, the recent progress in deformable convolution~\cite{Dai2017DeformableCN,Wang2019EDVRVR} has shown it is feasible to align two consecutive frames in the feature space by predicting the offsets of convolution kernels (\textit{i.e.}, the so-called dynamic kernels), where the learned offset maps are used to control which positions are sampled from the input features when performing the deformable convolution operations with the learned kernels.
Furthermore, the recent progress in deformable convolution~\cite{Dai2017DeformableCN,Wang2019EDVRVR} has shown it is feasible to align two consecutive frames in the feature space.
Based on the learned offset maps of convolution kernels (\textit{i.e.}, the so-called dynamic kernels), deformable convolution can decide which positions are sampled from the input feature. 
By using deformable convolution based on dynamic kernels, we can better cope with more complex non-rigid motion patterns between two consecutive frames, which can thus improve the motion compensation results and also alleviate the burden for the subsequent residual compression module.
Unfortunately, it is still a non-trivial task to seamlessly incorporate the feature space operations and deformable convolution into the existing hybrid deep video compression framework and train the whole network in an end-to-end fashion.

In our work, instead of following the existing works~\cite{lu2019dvc,hu2020imporving,agustsson2019generative,liu2020mlvc} to adopt the pixel-level operations as in the traditional video codecs, we propose a new feature-space video coding network (referred to as FVC) by reducing the spatial-temporal redundancy in the feature space.
Specifically, we first estimate motion information (\textit{i.e.}, the offset maps for convolution kernels in deformable convolution), based on the extracted representations from two consecutive frames. 
Then the offset maps are compressed by using the auto-encoder style network and the reconstructed offset maps will be used in the subsequent deformable convolution operation to generate the predicted feature for more accurate motion compensation. 
After that, we adopt another auto-encoder style network to compress the residual feature between the original input feature and the predicted feature. 
Finally, a multi-frame feature fusion module is proposed to combine a set of reference features from multiple previous frames for better frame reconstruction.

When compared with the state-of-the-art learning based video compression methods~\cite{lu2020content,hu2020imporving}, we perform all operations in the feature space for more accurate motion estimation and motion compensation, in which we can seamlessly incorporate deformable convolution into the hybrid deep video compression framework. 
As a result, we can alleviate the errors introduced by inaccurate pixel-level operations like motion estimation/compensation and achieve better video compression performance.
Our contributions are summarized as follows:

\begin{itemize}
  \item We propose a new learning based video compression framework, which performs all operations including motion estimation, motion compensation and residual compression in the feature space.

  \item For effective motion compensation in the feature space, we use deformable convolution to ``warp'' the reference feature from the previous reconstructed frame and more accurately predict the feature of the current frame, in which the corresponding offset maps are compressed by using the auto-encoder style network.
  
  \item We propose a new multi-frame feature fusion module based on the non-local attention mechanism, which combines the features from multiple previous frames for better reconstruction of the current frame.
  
  \item Our framework FVC achieves the state-of-the-art performance on four benchmark datasets including HEVC, UVG, VTL and MCL-JCV, which demonstrates the effectiveness of our proposed framework. 
  
\end{itemize}

\section{Related Work}

\subsection{Image Compression}
In the past decades, the traditional image compression methods like JPEG~\cite{wallace1992jpeg}, JPEG2000~\cite{taubman2002jpeg2000} and BPG~\cite{bellard2015bpg} have been proposed to reduce spatial redundancy, in which the hand-crafted techniques, such as DCT, are exploited for achieving high compression performance. Recently, the learning based image compression methods have attracted increasing attention. The RNN-based image compression methods~\cite{toderici2015variable, toderici2017full, johnston2018improved} are firstly introduced to progressively compress images. Other methods~\cite{balle2016end, balle2018variational, minnen2018joint, theis2017lossy} adopted the auto-encoder structure to first encode images as latent representations, and then decode the latent representations from the feature space to the pixel space, which have achieved the state-of-the-art performance for image compression.

\subsection{Video Compression}
A number of video compression standards~\cite{wiegand2003overview, sullivan2012overview} have been proposed in the past few years. Most methods follow the hybrid coding structure, where the motion compensation and residual coding techniques are used to reduce the redundancy in both spatial and temporal domains. Recently, learning based video compression \cite{wu2018video,lu2019dvc,rippel2019learned,habibian2019video,abdelaziz2019neural,liu2020mlvc,lu2020content,lu2020anendtoend, hu2020imporving, feng2020learned,Agustsson2020space,yang2020learning,cheng2019learning,xu2019non,lombardo2019deep,klopp2020utilising,yang2020learningh} has become a new research direction.
% Wu \etal~\cite{wu2018video} used the block-based motion estimation method and a newly proposed RNN-based approach to compress video sequences. 
Following the traditional hybrid video compression framework, Lu \etal~\cite{lu2019dvc} proposed the first end-to-end optimized video compression framework, in which all the key components in H.264/H.265 are replaced with deep neural networks.
% Rippel~\cite{rippel2019learned} proposed a framework to maintain the past states of motion and residual information for better compression. 
% Habibian~\cite{habibian2019video} directly proposed a 3D auto-encoder to compress the sequential data of video sequences.
% By formulating the motion compensation process as frame interpolation, Djelouah \etal~\cite{abdelaziz2019neural} proposed to compress the corresponding pixel-level motion information and feature-level residual information. 
Lin \etal~\cite{liu2020mlvc} used multiple frames in different modules to further remove the redundancy, while Hu \etal~\cite{hu2020imporving} proposed a resolution-adaptive optical flow compression method by automatically deciding the optimal resolution for each frame and each block.
% To improve video compression performance, Feng~\cite{feng2020learned} proposed a new method to compress feature-level residual information, which is then combined with pixel-level residual information. 

Most of the existing approaches~\cite{lu2019dvc,liu2020mlvc,hu2020imporving} have to estimate the pixel-level optical flow maps and compress the corresponding pixel-level residual information. However, it may be difficult to produce reliable pixel-level flow maps or residual information, which often degrades the compression performance of learning based video codecs. 
To address this problem, some recent works~\cite{abdelaziz2019neural,feng2020learned} proposed to calculate the feature-level residual maps for video compression.
In contrast to these works~\cite{abdelaziz2019neural,feng2020learned}, in our work, we propose to perform all operations in the feature space, which leads to better video compression performance.

\begin{figure*}[t]
% \begin{center}
\centering
\includegraphics[width=0.95\linewidth]{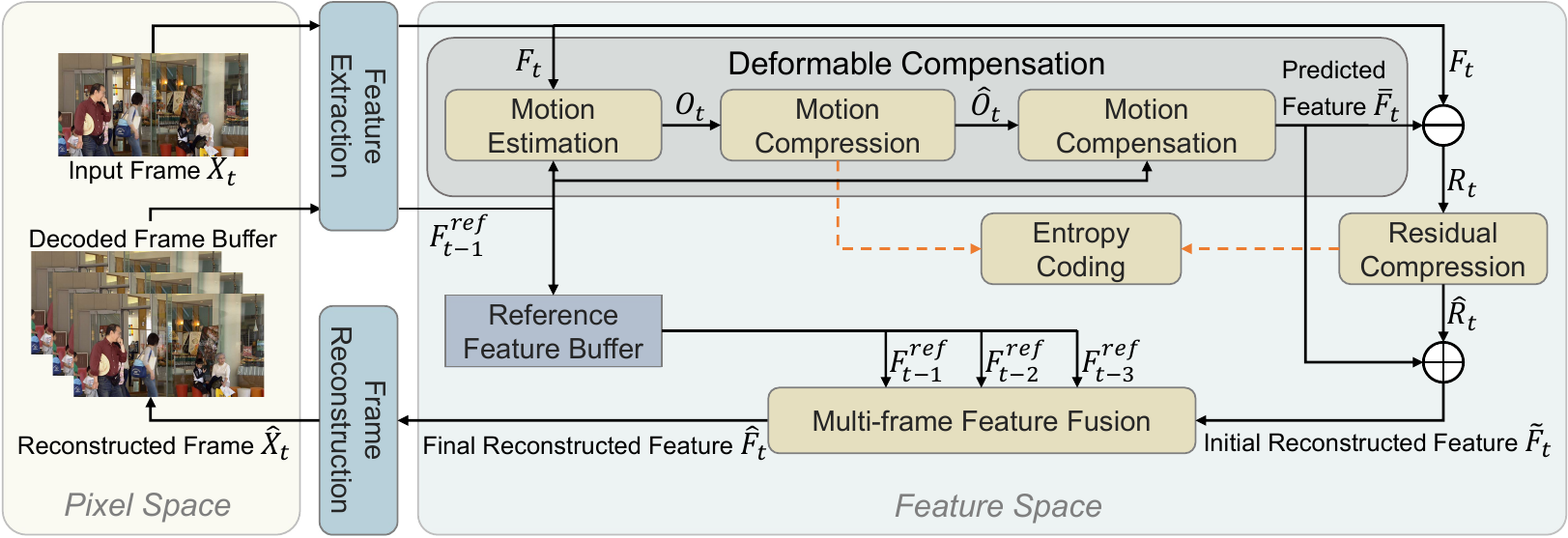}
% \end{center}
    \caption{Overview of our proposed video compression framework. Given an input frame $X_t$, we first encode the input frame to produce the input feature $F_t$. The \textbf{Deformable Compensation} module consists of three steps (\textit{i.e.}, motion estimation, motion compression and motion compensation). Specifically, the offset map $O_t$ between the features $F_t$ and $F_{t-1}^{ref}$ from the previous reconstructed frame is estimated and then used as motion information. The offset map is then compressed and the reconstructed offset map $\hat{O}_t$ is employed for motion compensation by using the deformable convolution operation. The residual feature $R_t$ between the input feature $F_t$ and the predicted feature $\Bar{F}_t$ is compressed in the \textbf{Residual Compression} module. Additionally, the \textbf{Multi-frame Feature Fusion} module is proposed to fuse the initial reconstructed feature $\tilde{F}_t$ and the features $F_{t-1}^{ref},F_{t-2}^{ref}$ and $F_{t-3}^{ref}$ from multiple previous frames to produce the final reconstructed feature $\hat{F_t}$. Finally, the reconstructed frame $\hat{X}_t$ is generated after going through the frame reconstruction module.}
\label{fig:overview}
\end{figure*}

\subsection{Deformable Convolution}
Recently, Dai \etal~\cite{Dai2017DeformableCN} proposed to use deformable convolution together with the learnt offset maps to enhance the modeling capability of convolution neural networks, which has achieved promising results in several video-related tasks like action recognition~\cite{Li2020SpatiotemporalD3} and video super-resolution~\cite{Tian2020TDANTA, Wang2019EDVRVR}. 
In contrast to these works, our work is the first to investigate how to employ deformable convolution in learning based video compression. Considering that there are multiple complex components in our hybrid video compression system, it is a non-trivial task to develop an end-to-end optimized video codec by seamlessly incorporating deformable convolution and simultaneously compressing the corresponding offset maps and other features.

\section{Methodology}

\subsection{Overview}

Let $X = \{X_1, X_2,...,X_{t-1}, X_t,...\}$ denote a video sequence, where $X_t$ is the original video frame at the current time step.
In our video compression system, the objective is to produce the high quality reconstructed frame $\hat{X}_t$ at any given bit-rate. As shown in Fig.~\ref{fig:overview}, all the modules in our proposed framework including deformable compensation, residual compression and multi-frame feature fusion are performed in the feature space. 
The overview architecture of our proposed framework is summarized as follows,

\begin{figure}[t]
% \begin{center}
\centering
\includegraphics[width=0.9\linewidth]{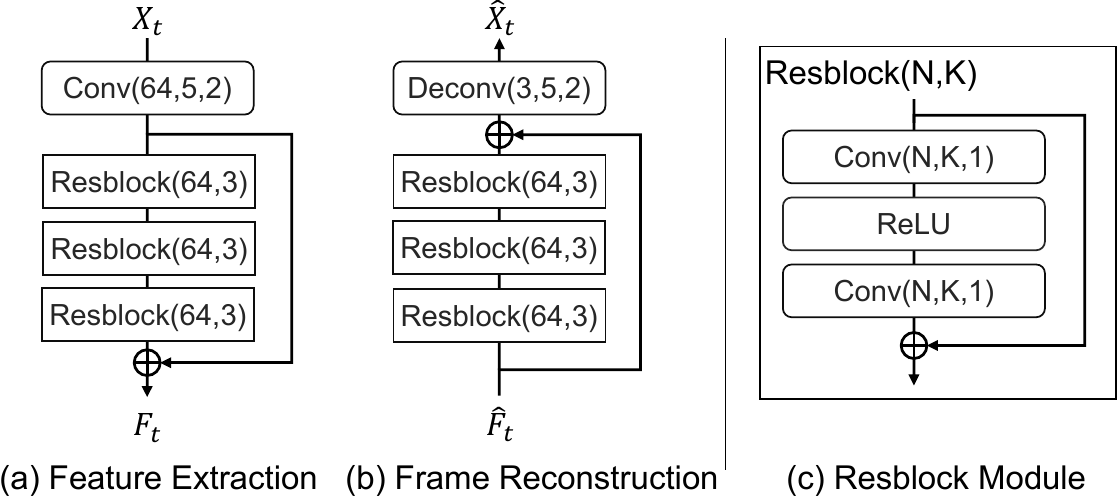}
% \end{center}
    \caption{The network structure of (a) our Feature Extraction module and (b) our Frame Reconstruction module with the details of each Resblock are shown in (c). ``Conv(N,K,S)" and ``Deconv(N,K,S)" represent the convolution and deconvolution operations with the output channel, the kernel size and the stride as $N$, $K\times K$ and $S$, respectively.}
\label{fig:edcoder}
\end{figure}

% \begin{figure}[t]
% \centering
%     \begin{minipage}[t]{}
%     \centering
%     \subfigure[Feature Extraction]{\includegraphics[width=0.9\linewidth]{figures/fcoder/encoder.pdf}\label{fig:encoder}}
%     \end{minipage}%
%     \begin{minipage}[t]{0.33\linewidth}
%     \centering
%     \subfigure[Frame Reconstruction]{\includegraphics[width=0.9\linewidth]{figures/fcoder/decoder.pdf}\label{fig:decoder}}
%     \end{minipage}%
%     \begin{minipage}[t]{0.34\linewidth}
%     \centering
%     \subfigure[Resblock module]{\includegraphics[width=0.9\linewidth]{figures/fcoder/resblock.pdf}\label{fig:resblock}}
%     \end{minipage}
% \label{fig:edcoder}
% \caption{The network structure of (a) our Feature Extraction module and (b) our Frame Reconstruction module with the details of each Resblock are shown in (c). ``Conv(N,K,S)" and ``Deconv(N,K,S)" represent the convolution and deconvolution operations with the output channel, the kernel size and the stride as $N$, $K\times K$ and $S$, respectively.}
% \end{figure}

\textbf{Feature Extraction.}
To produce the representations in the feature space, the original input frame $X_t$ and the previous reconstructed frame $\hat{X}_{t-1}$ are encoded as the feature representations $F_t$ and $F^{ref}_{t-1}$, respectively.
As shown in Fig.~\ref{fig:edcoder}(a), the feature extraction module uses a convolution layer with the stride of 2, which is then followed by several residual blocks to produce the feature representation.

\textbf{Deformable Compensation.}
This procedure consists of three steps: motion estimation, motion compression and motion compensation.
Specifically, based on $F_t$ and $F^{ref}_{t-1}$, we perform motion estimation by using a lightweight network, and the output offset map $O_t$ will be compressed by using the newly proposed motion compression network before being transmitted to the decoder side. Finally, given the reconstructed offset map $\hat{O}_t$ and the feature $F_{t-1}^{ref}$, we can generate the predicted feature $\bar{F}_{t}$ by using deformable convolution in the motion compensation procedure. More details can be found in Section~\ref{subsec:compensation}.

\textbf{Residual Compression.}
The residual feature $R_t$ between the input feature $F_t$ and the predicted feature $\Bar{F}_t$ will be compressed by using an auto-encoder style network for better reconstruction. After adding the reconstructed residual feature $\hat{R}_t$ back to the predicted feature $\Bar{F}_t$, we produce the initial reconstructed feature representation $\Tilde{F}_t$.

\textbf{Multi-frame Feature Fusion.}
In this procedure, the extracted feature representations $F^{ref}_{t-1}$, $F^{ref}_{t-2}$ and $F^{ref}_{t-3}$ from three previous reconstructed frames as well as the initial reconstructed feature representation $\Tilde{F}_t$ are fused to produce the final reconstructed feature $\hat{F}_t$.
More details will be introduced in Section~\ref{subsec:fusion}.

\textbf{Frame Reconstruction.}
As shown in Fig.~\ref{fig:edcoder}(b), the feature decoder with several residual blocks and a deconvolution layer will transform the final reconstructed feature $\hat{F}_t$ to the reconstructed frame $\hat{X_t}$.

\textbf{Entropy Coding.}
The quantized features from the motion compression and residual compression modules will be transformed into the bit-streams by performing entropy coding. During the training process, we use a bit estimation network to estimate the number of bits. More details will be described in section~\ref{subsec:compress}.

\subsection{Deformable Compensation}
\label{subsec:compensation}

The previous video compression frameworks~\cite{lu2019dvc, abdelaziz2019neural, hu2020imporving} rely on the optical flow estimation network and the pixel-level motion compensation network, which may lead to inaccurate frame prediction results and thus bring extra redundancy to the subsequent residual compression module. 
To this end, we perform motion compensation in the feature space and generate the predicted features by using the deformable convolution operation.

% the deformable convolution operation~\cite{Dai2017DeformableCN} by using multiple offsets for each spatial location of the encoded feature map, which makes motion compensation more robust.

\begin{figure}[t]
% \begin{center}
\centering
\includegraphics[width=0.9\linewidth]{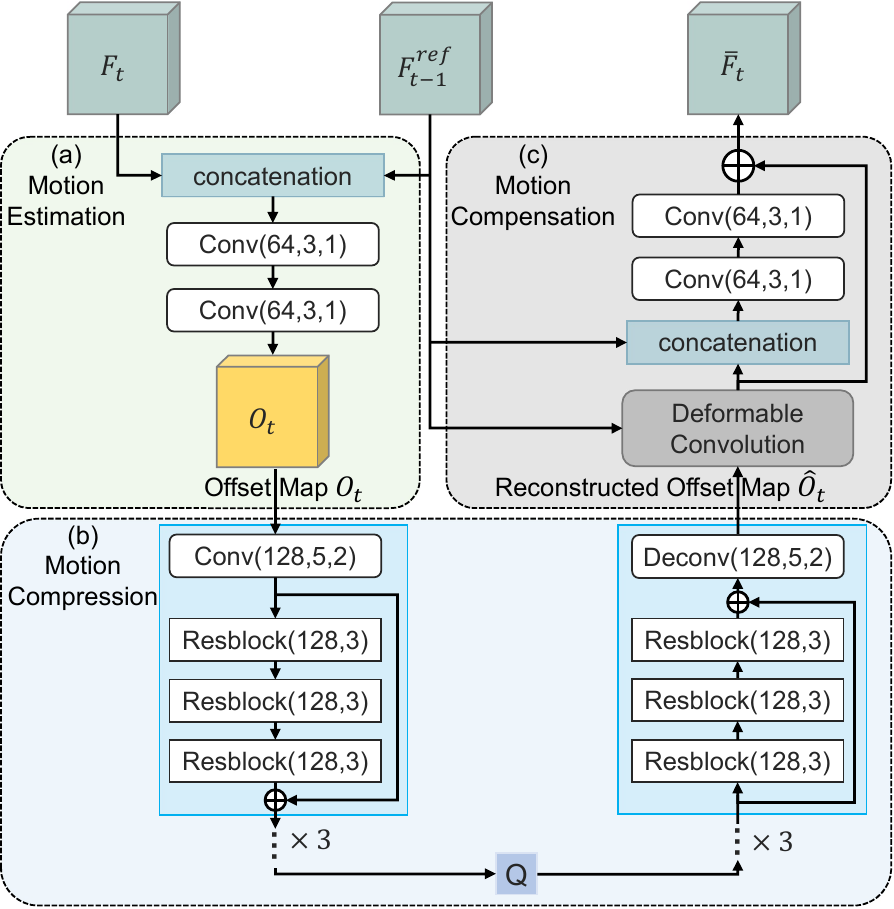}
% \end{center}
    \caption{The network structure of our \textbf{Deformable Compensation} module, which consists of (a) the motion estimation submodule, (b) the motion compression submodule and (c) the motion compensation submodule for reducing redundancy.}
\label{fig:compensation}
\end{figure}

Given the features $F^{ref}_{t-1}$ and $F_t$ respectively from the previous reconstructed frame and the current frame, the deformable compensation procedure aims at generating the predicted feature $\bar{F}_t$ at the current time step. The whole network architecture is shown in Fig.~\ref{fig:compensation}.
Specifically, we first produce the deformable offset map $O_t$ by feeding  $F^{ref}_{t-1}$ and $F_t$ into a 2-layer motion estimation network. 
Inspired by~\cite{feng2020learned}, we propose an auto-encoder style network to compress this offset map $O_t$, where both the encoder and the decoder consist of a set of Resblocks~\cite{He2016DeepRL}. The offset map $O_t$ is transformed to the latent space through the encoder and then quantized. After that, the decoder will convert the quantized latent representations back to the reconstructed offset map $\hat{O}_t$.
% The details of our motion compression network will be provided in the supplementary material.

Finally, in the motion compensation procedure, the deformable convolution layer takes $F^{ref}_{t-1}$ as the input and then performs the deformable convolution operation by using the corresponding filters with the aid of $\hat{O}_t$. 
In this way, we can better compress videos with complex non-rigid motion patterns by using the learned dynamic offset kernels and produce a more accurate warped feature.
To generate a more accurate predicted feature, we further refine the output feature from the deformable convolution layer by using two convolution layers and eventually produce the final predicted feature $\bar{F}_t$.

\begin{figure}[t]
\centering
\includegraphics[width=0.7\linewidth]{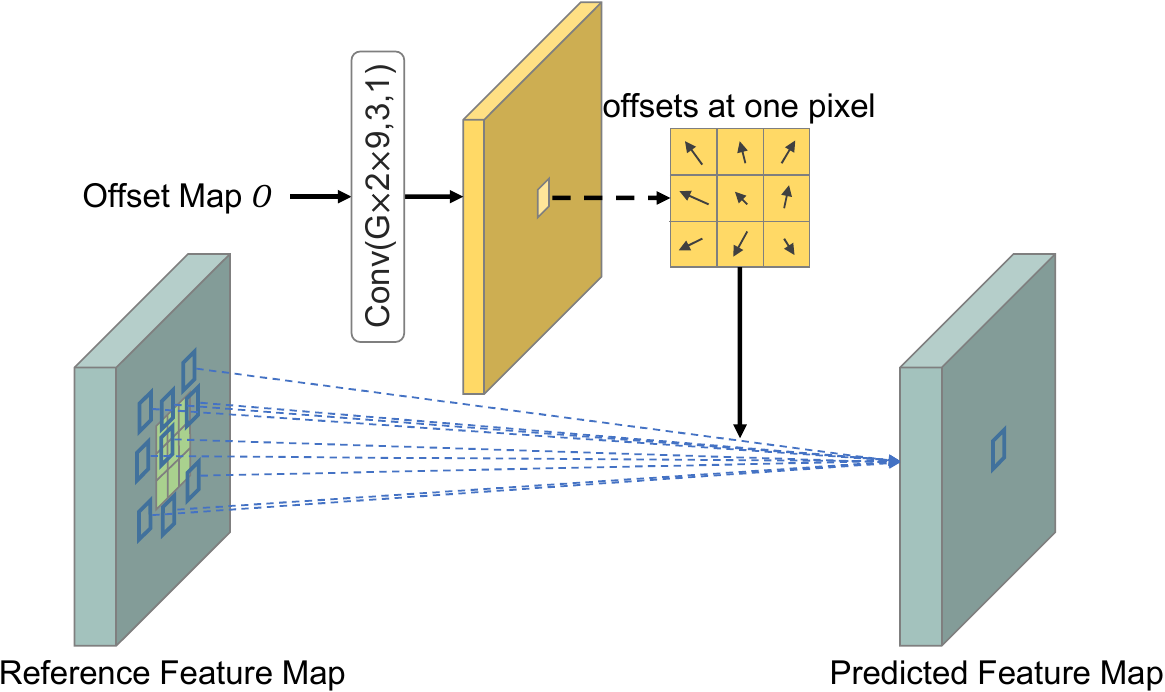}
    \caption{Illustration of Deformable Convolution. For the output channel of the convolution layer, ``G", ``2", ``9" denote the channel group ($G=8$), two directions (\textit{i.e.}, the horizontal and vertical directions) of the offset map and the size of each kernel ($3\times3$), respectively.}%The output channel is $G\times2\times9$, in which ``$G$" represents the channel group ($G=8$), ``$2$" represents two directions (\textit{i.e.}, horizontal and vertical directions) of the offset and ``$9$" denotes the size of each kernel is $3\times3$.}
\label{fig:deformable}
\end{figure}

\textbf{Deformable Convolution.} 
The network structure of our deformable convolution layer~\cite{Dai2017DeformableCN} is shown in Fig.~\ref{fig:deformable}. Given the reference feature map and the corresponding offset map $O$, we aim to generate the predicted feature. 
For each kernel, the corresponding offsets are used to control the sampling location in the reference feature map, and then the feature values from the corresponding locations in the reference feature map are fused to generate one feature value in the predicted feature map.
When compared with the pixel-level warping operation based on motion compensation as used in the previous video compression approach~\cite{lu2019dvc}, our deformable compensation module provides more flexibility by using different sampling locations, which can better cope with complex non-rigid motion patterns and improve the compensation performance. 
In our implementation, we divide the channels of the reference feature into G groups (G=8 in this work) and use a shared offset map for each group of channels to better learn the offset maps and improve the deformable compensation results. %Please refer to the supplementary material for more details.

\subsection{Multi-frame Feature Fusion}
\label{subsec:fusion}

\begin{figure}[t]
\centering

\subfigure[The overall network structure for multi-frame feature fusion.]{\includegraphics[width=\linewidth]{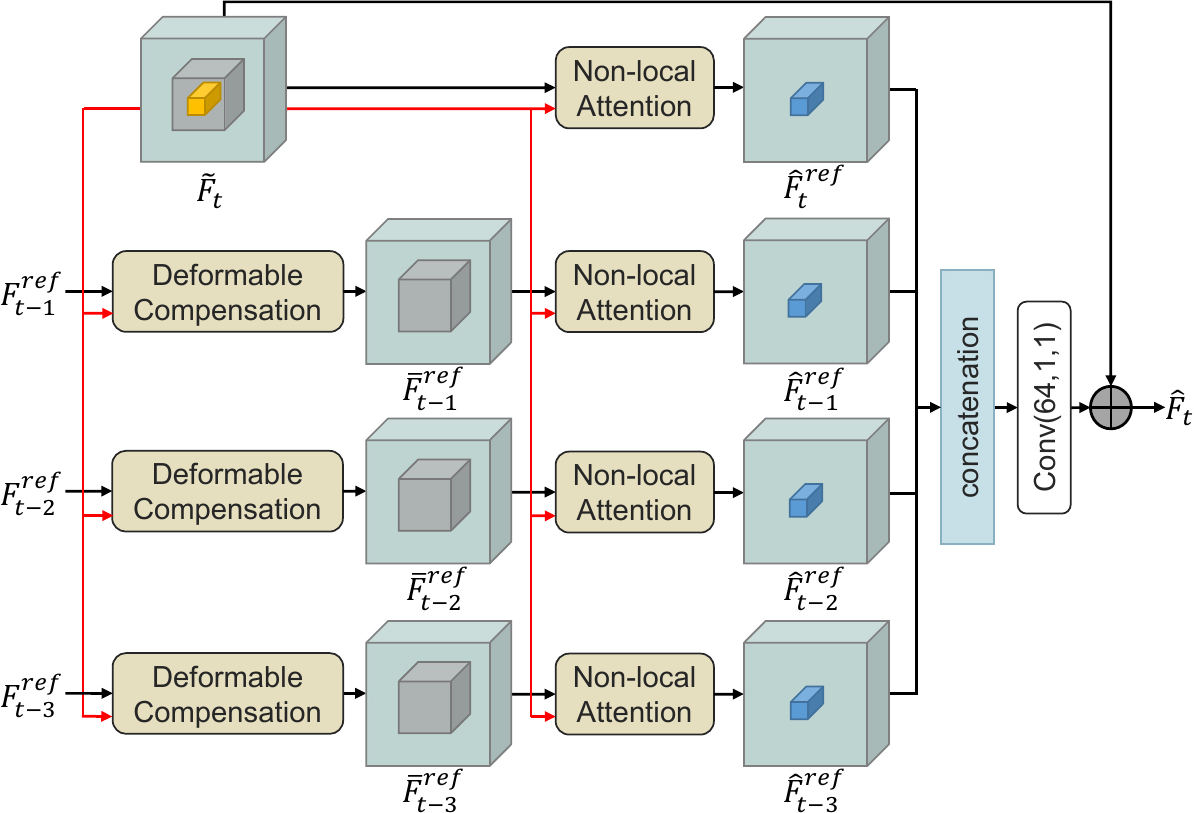}\label{subfig:non-fusion}}
\subfigure[Non-local attention mechanism.]{\includegraphics[width=0.5\linewidth]{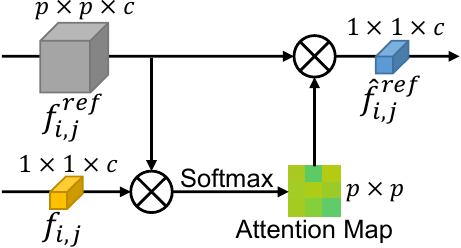}\label{subfig:nonatt}}

\centering
    \caption{Illustration of our multi-frame feature fusion module.}% $\tilde{F}$ denotes the initial reconstructed feature of the current frame and $F_{t-1}^{ref}, F_{t-2}^{ref}, F_{t-3}^{ref}$ denotes the feature representations from previous reconstructed frames. $\hat{F}_t$ represents the final reconstructed feature of the current frame.}
\label{fig:fusion}
\end{figure}

After the deformable compensation and residual compression procedures, we produce the initial reconstructed feature $\tilde{F}_t$ by combining the predicted feature $\bar{F}_t$ and the reconstructed residual feature $\hat{R}_t$.
As shown in Fig.~\ref{fig:overview}, to generate more accurate final reconstructed feature $\hat{F}_t$,  a multi-frame feature fusion module is proposed to exploit temporal context information from the feature representations $F^{ref}_{t-1}$, $F^{ref}_{t-2}$, $F^{ref}_{t-3}$ of multiple previous frames.

The detailed network architecture is shown in Fig.~\ref{subfig:non-fusion}.
Specifically, we first produce the predicted feature representations $\bar{F}^{ref}_{t-1}$, $\bar{F}^{ref}_{t-2}$ and $\bar{F}^{ref}_{t-3}$ by using the shared feature-space deformable compensation module. 
The non-local attention mechanism is then used to refine the predicted feature representation based on the similarity between the initial reconstructed feature $\tilde{F}_t$ and the predicted representations $\bar{F}^{ref}_{t-1}$, $\bar{F}^{ref}_{t-2}$ and $\bar{F}^{ref}_{t-3}$. 
It is worth mentioning that we also refine $\tilde{F}_t$ itself by using the non-local attention mechanism based on the so-called self-attention mechanism. 
Finally, we concatenate all these refined feature representations and use another convolution layer to generate the final reconstructed feature $\hat{F}_t$. 

The network architecture of the deformable compensation module is the same as that in section~\ref{subsec:compensation}.
As for the non-local attention block, we provide the detailed architecture in Fig.~\ref{subfig:nonatt}.  Here we take $\bar{F}^{ref}_{t-1}$ as an example to describe how to produce the refined feature $\hat{F}_{t-1}^{ref}$ after performing the attention operation guided by $\tilde{F}_t$, in which we assume the dimension of these two features $\bar{F}^{ref}_{t-1}$ and $\tilde{F}_t$ is $w \times h \times c$. For each spatial feature vector $f_{i,j}$ from $\tilde{F}_t$ at the location $(i,j)$, whose size is $1 \times 1 \times c$, we find a collocated patch $f^{ref}_{i,j}$ with the size of $p \times p \times c$ from $\bar{F}^{ref}_{t-1}$. Then we calculate the similarity between $f_{i,j}$ and $f^{ref}_{i,j}$ by performing the convolution operation along the channel dimension and then produce the corresponding attention map after performing the softmax operation, which is further used to re-weight all spatial positions in $f^{ref}_{i,j}$. After that, we generate the refined feature vector $\hat{f}^{ref}_{i,j}$ at the location $(i,j)$. We repeat this procedure for each spatial location in $\tilde{F}_t$ and then produce the refined feature $\hat{F}^{ref}_{t-1}$. Finally, we concatenate all refined features $\hat{F}_{t-3}^{ref}$, $\hat{F}_{t-2}^{ref}$, $\hat{F}_{t-1}^{ref}$ and $\hat{F}_{t}^{ref}$ and perform the convolution operation to fuse these refined features and produce the final reconstructed feature $\hat{F}_t$ by adding the initial reconstructed feature $\tilde{F}_t$ back.

\subsection{Residual Compression and Other Details}
\label{subsec:compress}

As shown in Fig.~\ref{fig:overview}, in addition to offset map compression, we also need to compress the residual feature map. To simplify the whole system, we adopt the same network architecture as that used for the offset feature map compression (see Fig.~\ref{fig:compensation}(b)).

For the whole learning based video compression framework, we use the bit-rate estimation network to generate the bit-rate in the training stage. In our implementation, we employ the hyperprior entropy model in \cite{minnen2018joint} for accurate bit-rate estimation. To reduce the computational complexity, the time-consuming auto-regressive model is not used in our approach.
%Please refer to the supplementary material for more details about the network architecture.

To optimize the whole model in an end-to-end manner, it is also required to design a differentiable quantization operation. In our approach, we follow the method in \cite{balle2018variational} and approximate the quantization operation by adding the uniform noise in the training stage.  In the inference stage, we directly use the \textit{rounding} operation.

\subsection{Loss Function}
% \vspace{-1mm}
\label{subsec:training}
In our proposed framework, we optimize the following Rate Distortion~(RD) trade-off:
\begin{equation}
\vspace{-1mm}
    RD = R + \lambda D =  R_o + R_r + \lambda d(X_t, \hat{X}_t),
\vspace{-1mm}
\end{equation}
where $R_o$ and $R_r$ denote the numbers of bits used to encode the offset map $O_t$ and the residual feature map $R_t$, respectively. $d(X_t, \hat{X}_t)$ denotes the distortion between the input frame $X_t$ and the reconstructed frame $\hat{X}_t$, where $d(\cdot)$ represents the mean square error or MS-SSIM~\cite{wang2003multiscale}. $\lambda$ is a hyper parameter used to control the rate-distortion trade-off.

\vspace{-1mm}

\section{Experiments}
\vspace{-1mm}

% \begin{figure*}[t]
%   \centering
%   \begin{minipage}[c]{0.25\textwidth}
%   \centering
%     \includegraphics[width=\textwidth]{figures/performance/HEVCClass_C_psnr.pdf}
%   \end{minipage}%
%   \begin{minipage}[c]{0.25\textwidth}
%   \centering
%     \includegraphics[width=\textwidth]{figures/performance/HEVCClass_D_psnr.pdf}
%   \end{minipage}%
%   \begin{minipage}[c]{0.25\textwidth}
%   \centering
%     \includegraphics[width=\textwidth]{figures/performance/UVG_psnr.pdf}
%   \end{minipage}%
%   \begin{minipage}[c]{0.25\textwidth}
%   \centering
%     \includegraphics[width=\textwidth]{figures/performance/VTL_psnr.pdf}
%   \end{minipage}
%   \begin{minipage}[c]{0.25\textwidth}
%   \centering
%     \includegraphics[width=\textwidth]{figures/performance/HEVCClass_C_msssim.pdf}
%   \end{minipage}%
%   \begin{minipage}[c]{0.25\textwidth}
%   \centering
%     \includegraphics[width=\textwidth]{figures/performance/HEVCClass_D_msssim.pdf}
%   \end{minipage}% 
%   \begin{minipage}[c]{0.25\textwidth}
%   \centering
%     \includegraphics[width=\textwidth]{figures/performance/UVG_msssim.pdf}
%   \end{minipage}%
%   \begin{minipage}[c]{0.25\textwidth}
%   \centering
%     \includegraphics[width=\textwidth]{figures/performance/VTL_msssim.pdf}
%   \end{minipage}
%     \caption{The experimental results on the HEVC Class C, Class D, UVG and VTL datasets. When using MS-SSIM for performance evaluation, we additionally perform the fine-tune operation in our FVC$^\#$ by using MS-SSIM as the distortion loss.}
%   \label{fig:result}
% \end{figure*}

% \begin{comment}
\begin{figure*}[t]
  \centering
  \begin{minipage}[c]{0.25\textwidth}
  \centering
    \includegraphics[width=\textwidth]{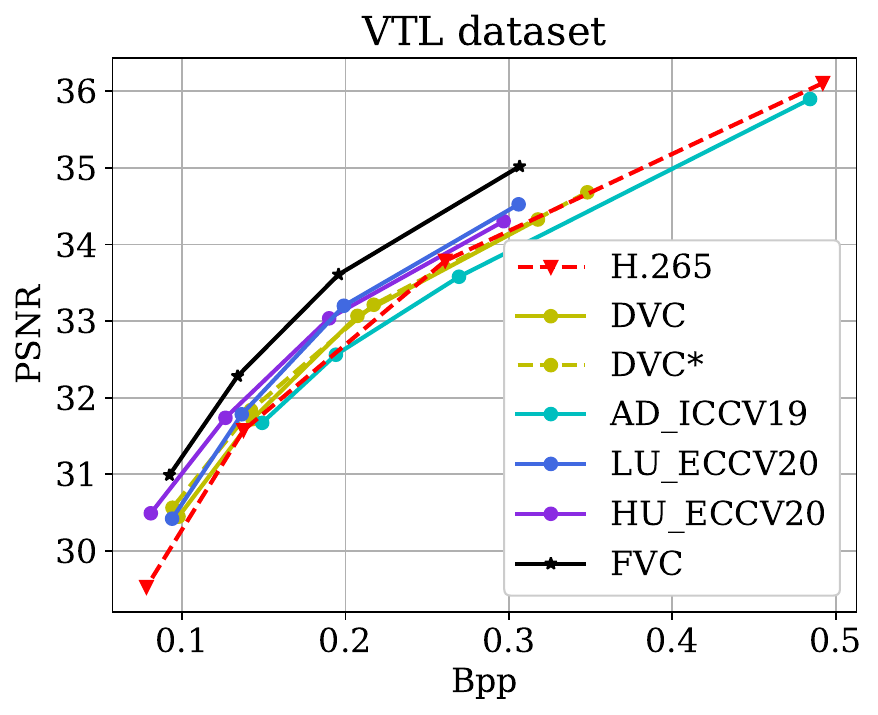}
  \end{minipage}%
  \begin{minipage}[c]{0.25\textwidth}
  \centering
    \includegraphics[width=\textwidth]{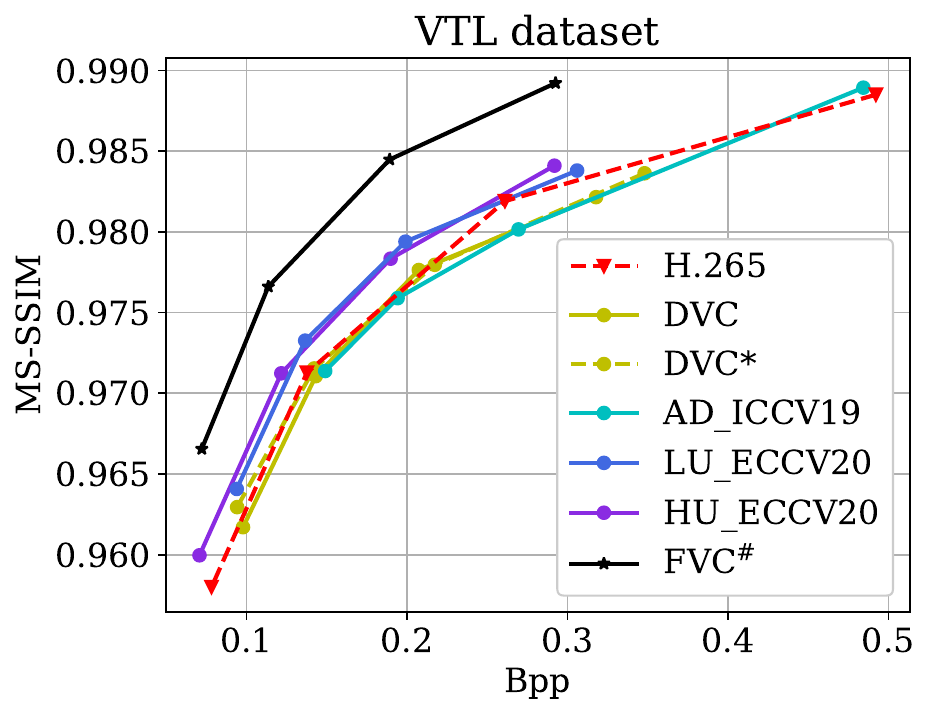}
  \end{minipage}%
  \begin{minipage}[c]{0.25\textwidth}
  \centering
    \includegraphics[width=\textwidth]{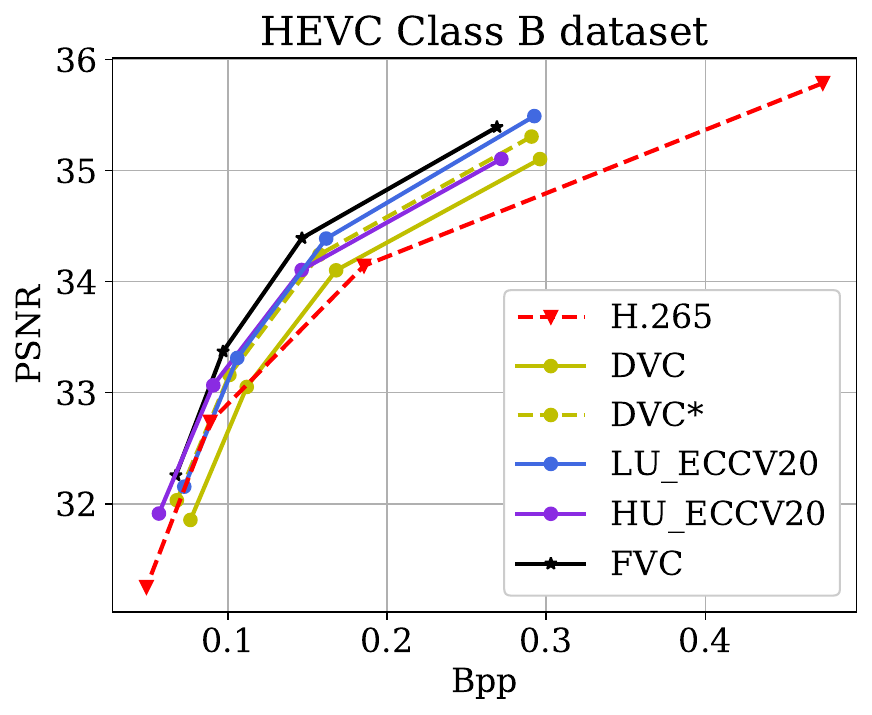}
  \end{minipage}%
  \begin{minipage}[c]{0.25\textwidth}
  \centering
    \includegraphics[width=\textwidth]{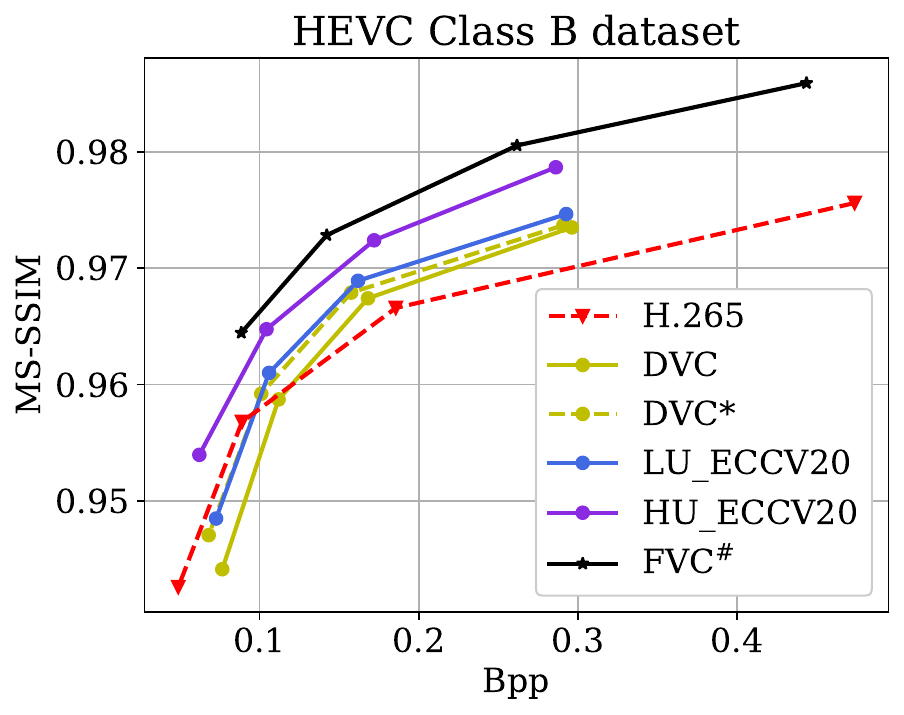}
  \end{minipage}
  \begin{minipage}[c]{0.25\textwidth}
  \centering
    \includegraphics[width=\textwidth]{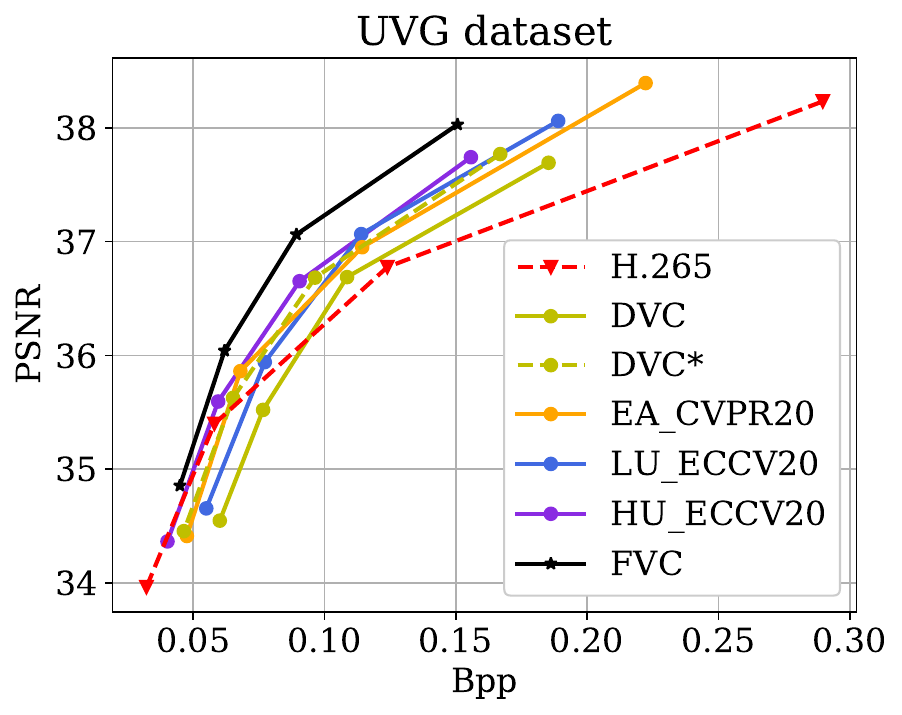}
  \end{minipage}%
  \begin{minipage}[c]{0.25\textwidth}
  \centering
    \includegraphics[width=\textwidth]{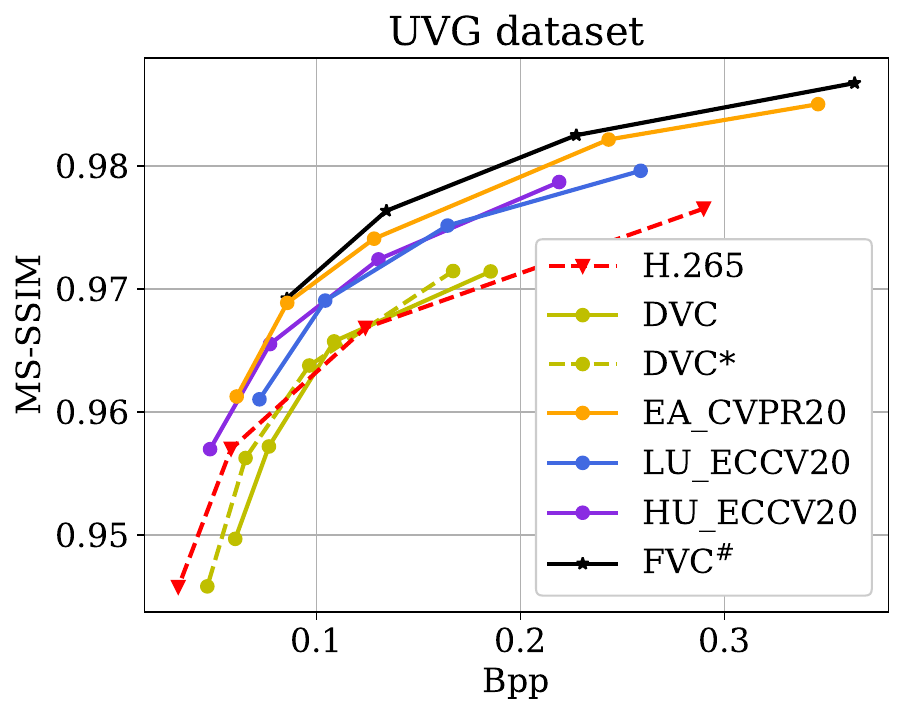}
  \end{minipage}%
  \begin{minipage}[c]{0.25\textwidth}
  \centering
    \includegraphics[width=\textwidth]{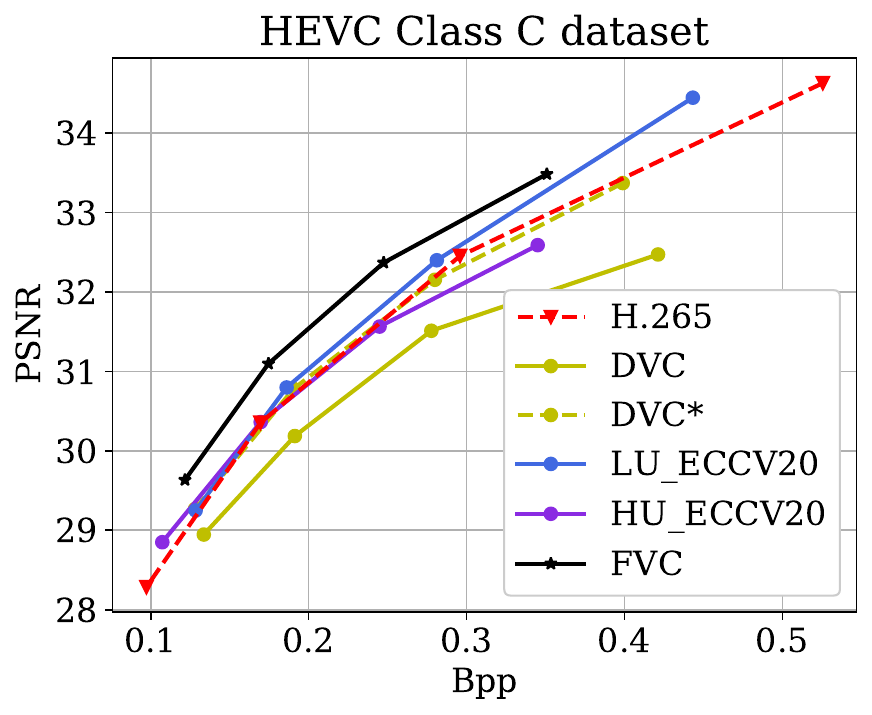}
  \end{minipage}%
  \begin{minipage}[c]{0.25\textwidth}
  \centering
    \includegraphics[width=\textwidth]{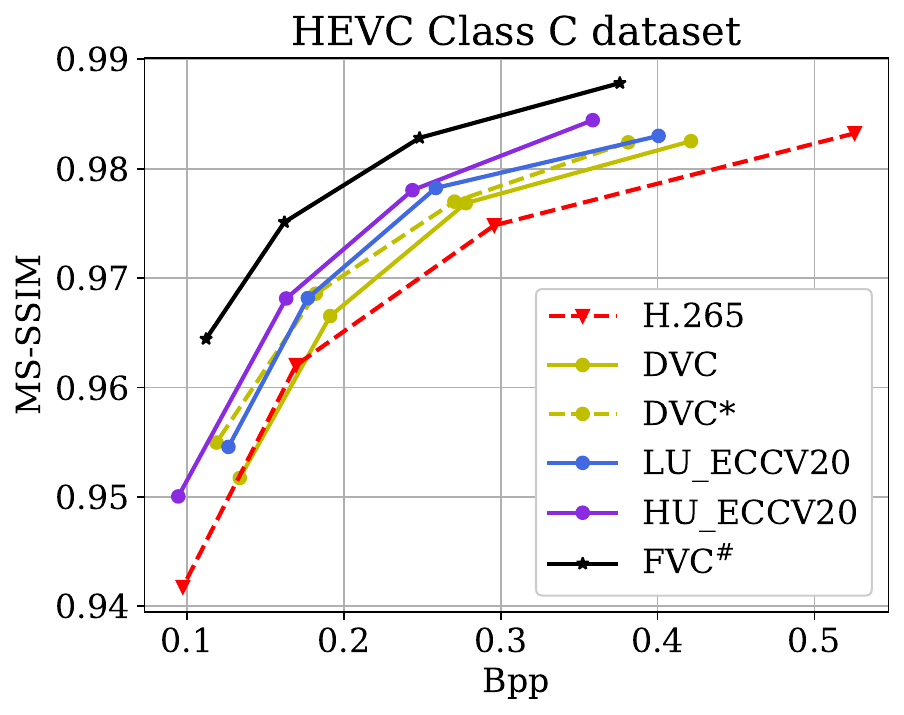}
  \end{minipage} 
  \begin{minipage}[c]{0.25\textwidth}
  \centering
    \includegraphics[width=\textwidth]{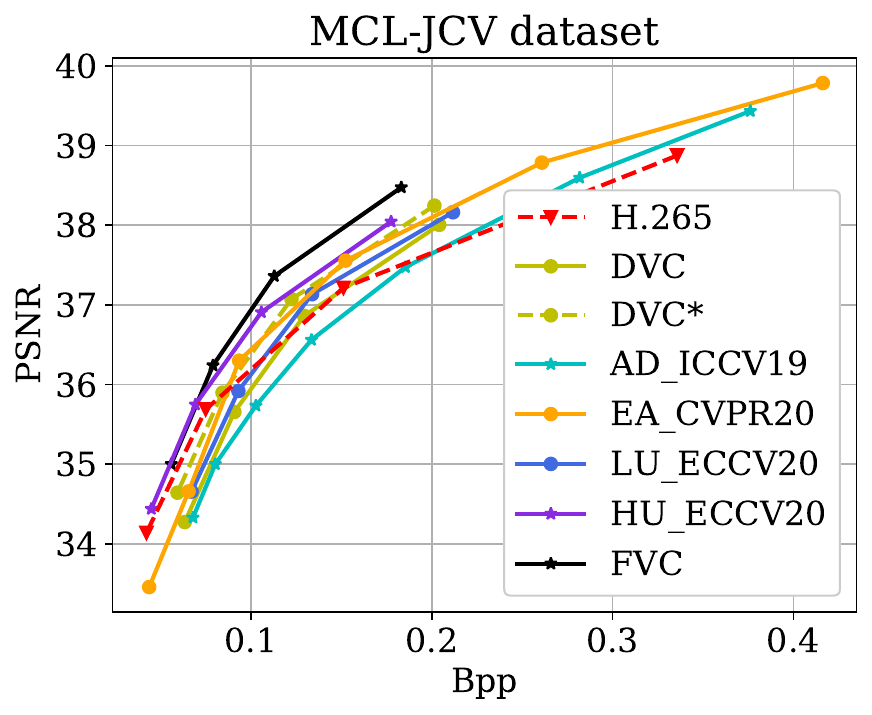}
  \end{minipage}%
  \begin{minipage}[c]{0.25\textwidth}
  \centering
    \includegraphics[width=\textwidth]{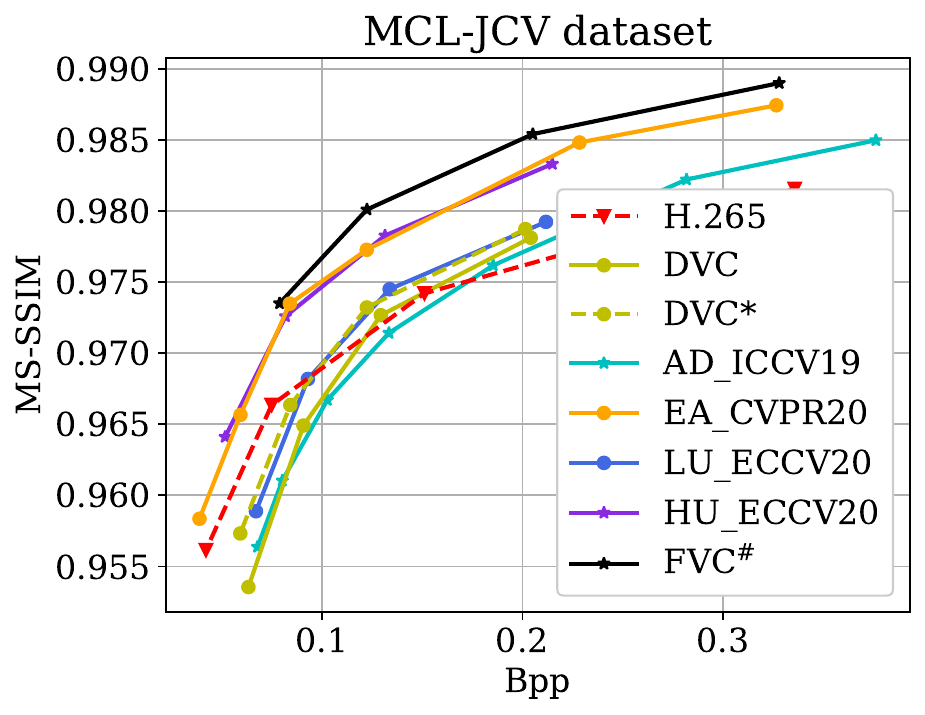}
  \end{minipage}%
  \begin{minipage}[c]{0.25\textwidth}
  \centering
    \includegraphics[width=\textwidth]{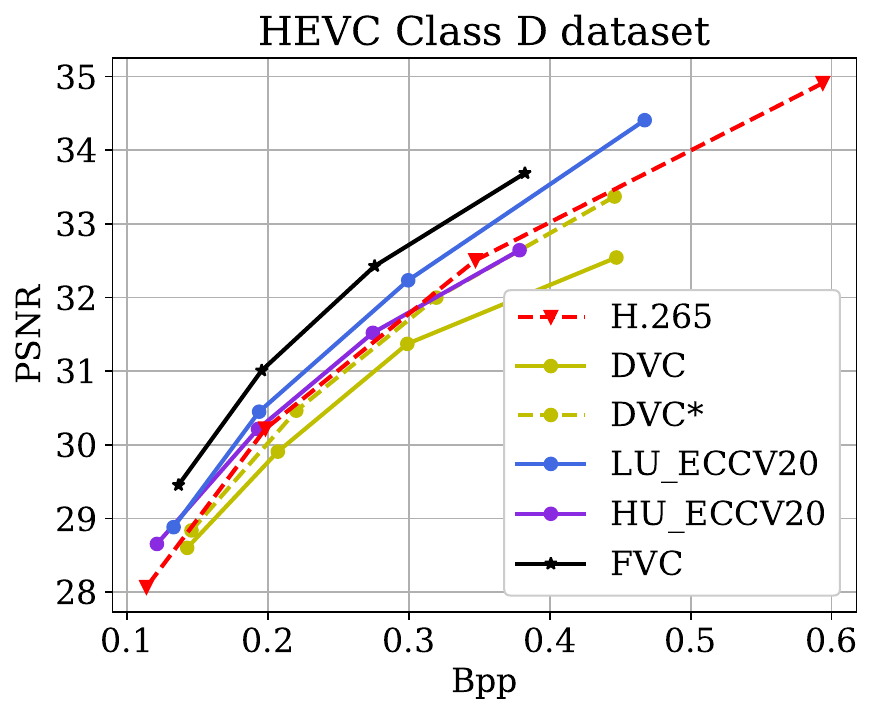}
  \end{minipage}%
  \begin{minipage}[c]{0.25\textwidth}
  \centering
    \includegraphics[width=\textwidth]{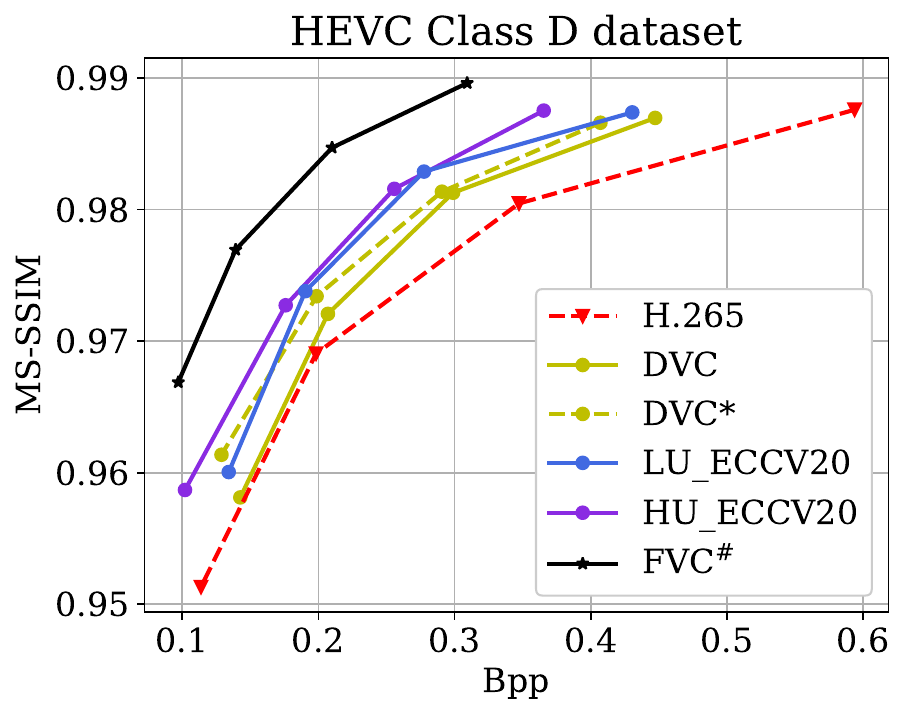}
  \end{minipage}
    \caption{The experimental results on the VTL, UVG, MCL-JCV and HEVC Class B, Class C, Class D datasets. When using MS-SSIM for performance evaluation, we additionally perform the fine-tune operation in our FVC$^\#$ by using MS-SSIM as the distortion loss.}
  \label{fig:result}
\end{figure*}

% \end{comment}

\subsection{Experimental Setup}
\textbf{Training Dataset.}
We use the Vimeo-90K dataset~\cite{xue2019video} in the training stage, which contains 89,800 video clips with each video having 7 frames with the resolution of $448 \times 256$. We randomly crop the video sequences into the resolution of $256\times256$ before the training process.

\textbf{Testing Datasets.}
To evaluate the performance of our FVC, we use the video sequences in four datasets including the HEVC~\cite{sullivan2012overview}, UVG~\cite{UVGdataset}, MCL-JCV~\cite{wang2016mcl} and VTL~\cite{VTLdataset}. The HEVC datasets contain 16 videos in Class B, C, D and E with different resolutions from $416\times240$ to $1920\times1080$. The UVG dataset contains seven high frame rate videos with the resolution of $1920\times1080$, in which the difference between the neighboring frames is small. The MCL-JCV dataset is widely used for video quality evaluation, which consists of thirty 1080p video sequences. For the VTL dataset, we use the first 300 frames from the high-resolution video sequences with the resolution of $352\times 288$ for performance evaluation.

\textbf{Evaluation Metrics.}
We use bpp (bit per pixel) to measure the average number of bits used for motion coding and residual coding for one pixel in each frame. We use PSNR and MS-SSIM\cite{wang2003multiscale} to evaluate the distortion between the reconstructed frame and the ground-truth frame and the PSNR/MS-SSIM of each video sequence is produced by calculating the average PSNR/MS-SSIM over all reconstructed frames. 
% PSNR is widely used for image quality evaluation and MS-SSIM has been adopted in recent works for subjective visual quality evaluation.

\textbf{Implementation Details.}
We train four models with different $\lambda$ values ($\lambda$=256, 512, 1024, 2048).  A two-stage training scheme is used to train our model. In the first training stage, we train the model without using the multi-frame feature fusion module for 2,000,000 steps at the learning rate of 5e-5. In the second stage, the multi-frame feature fusion module is included and we first train the overall framework with the learning rate of 5e-5 for 400,000 steps, and then optimize the model with the reduced learning rate of 5e-6 for 100,000 steps. When using MS-SSIM for performance evaluation, we further fine-tune the model from Stage 2 for 80,000 steps by using MS-SSIM as the distortion loss to achieve better MS-SSIM results. Our method is implemented based on PyTorch with CUDA support. All the experiments are conducted on the machine with a single NVIDIA 2080TI GPU (11GB memory). We set the batch size as 4 and use the Adam optimizer~\cite{kingma2014adam}. It takes about 4 days, 3 days and 12 hours for the first training stage, the second training stage and the fine-tuning stage, respectively.

% In our deformable convolution layer, given the input feature and the corresponding offset map, the offset map will first go through a convolution layer to generate the offset for each kernel/filter, which represents motion information. The size of the output channel is $G\times2\times9$, in which ``$G$" represents the channel group ($G=8$), ``$2$" represents two directions (\textit{i.e.}, the horizontal and vertical directions) of the offset and ``$9$" denotes the size of each kernel is $3\times3$.
% Namely, we have a shared offset map for each group of channels. For each kernel, the corresponding offset is used to control the sampling location in the input feature map.
% In our deformable convolution layer, we divide the channels of the input feature into 8 groups and use a shared offset map for each group of channels to speed up the deformable convolution operation. Please refer to the supplementary material for more details.

\subsection{Experimental Results}

\begin{table}[!t]
\centering
\small
    % \begin{center}
            \caption{BDBR(\%) results of DVC~\cite{lu2019dvc}, EA~\cite{Agustsson2020space}, LU~\cite{lu2020content}, HU~\cite{hu2020imporving} and our proposed method FVC when compared with H.265~\cite{sullivan2012overview} on different datasets. Negative values in BDBR indicate bit-rate savings while positive values indicate more bit-rate costs.}\setlength{\tabcolsep}{2mm}{
    \begin{tabular}{|l|c|c|c|c|c|c|c|}
    \hline
        & DVC & EA & LU & HU & FVC\\
    \hline
        HEVC Class B & 2.97   &   -    & -15.92 & -14.91          & \textbf{-23.75}\\
    \hline
        HEVC Class C & 20.65  &   -    & -3.78  & 1.76            & \textbf{-14.18}\\
    \hline
        HEVC Class D & 14.08  &   -    & -8.29  & -1.77           & \textbf{-18.39}\\
    \hline
        UVG          & 8.45   &  -9.75 & -7.34  & -13.27          &\textbf{-28.71}\\
    \hline
        VTL          & -10.92 &   -    & -16.85 & -20.17          &\textbf{-28.10}\\
    \hline
        MCL-JCV      & 13.94  &  -1.52 &  4.75  & -13.71          &\textbf{-22.48}\\
    \hline
    \end{tabular}}
    \label{tab:BDBR}
\end{table}

\textbf{Settings of the Baseline Methods.}
In this section, we provide the experimental results to demonstrate the effectiveness of our proposed method FVC on four benchmark datasets HEVC~\cite{sullivan2012overview}, UVG~\cite{UVGdataset}, VTL~\cite{VTLdataset} and MCL-JCV~\cite{wang2016mcl}. We compare our method with other state-of-the-art methods including the traditional methods~\cite{wiegand2003overview,sullivan2012overview} and recently proposed learning based methods (DVC~\cite{lu2019dvc}, AD\_ICCV19~\cite{abdelaziz2019neural}, EA\_CVPR20~\cite{Agustsson2020space}, LU\_ECCV20~\cite{lu2020content} and HU\_ECCV20~\cite{hu2020imporving}). For the traditional method H.265~\cite{sullivan2012overview}, we follow the command line in \cite{hu2020imporving} and use FFmpeg with the \textit{default} mode. % and more details can be found in our supplementary material. 
For fair comparison, we additionally report the results of ``DVC*", which is an enhanced version of DVC by using the same motion and residual encoder/decoder modules as those in our deformable compensation module (see Fig.~\ref{fig:compensation}).
Following the previous methods~\cite{lu2019dvc,hu2020imporving,lu2020content}, we set the GoP size as 10 for the HEVC datasets and 12 for other datasets. We use the same I-frame compression method as in H.265 to reconstruct I-frame. For our ``FVC$^\#$", we further perform the fine-tune operation according to the MS-SSIM based rate-distortion trade-off.

\textbf{Results.}
In Table~\ref{tab:BDBR}, we report the BDBR~\cite{bjontegaard2001calculation} results of different video compression methods when compared with H.265 on the HEVC, UVG, VTL and MCL-JCV datasets.
Specifically, our approach saves more than 18\% bit rate in terms of the overall results on all benchmark datasets. It is obvious that our approach outperforms DVC and other recent state-of-the-art learning based video compression approaches. 
For example, when compared with H.265, our proposed FVC saves 18.39\% bit-rate on the HEVC Class D dataset while the corresponding bit-rate savings are 8.29\% and 1.77\% for the recent approaches LU\_ECCV20~\cite{lu2020content} and HU\_ECCV20~\cite{hu2020imporving}.

We provide the RD curves of different compression methods in Fig.~\ref{fig:result}, it is noted that our method outperforms the baseline method DVC~\cite{lu2019dvc} and the enhanced version DVC* by a large margin on all datasets. We would like to highlight that DVC* directly compresses the pixel-level optical flow maps and the pixel-level residual maps by using the same compression network as our proposed method FVC, so the performance improvement clearly demonstrates the effectiveness of our proposed method. When compared with the traditional method H.265, our method achieves better results in terms of PSNRs at all bit-rates. 
Our method also outperforms all other baseline methods in terms of MS-SSIM.

\subsection{Ablation Study and Analysis}
% \vspace{-1mm}

\begin{figure}[t]
% \begin{center}
\centering
    \begin{minipage}[t]{0.5\linewidth}
    \centering
    \subfigure[]{\includegraphics[height=1.23in]{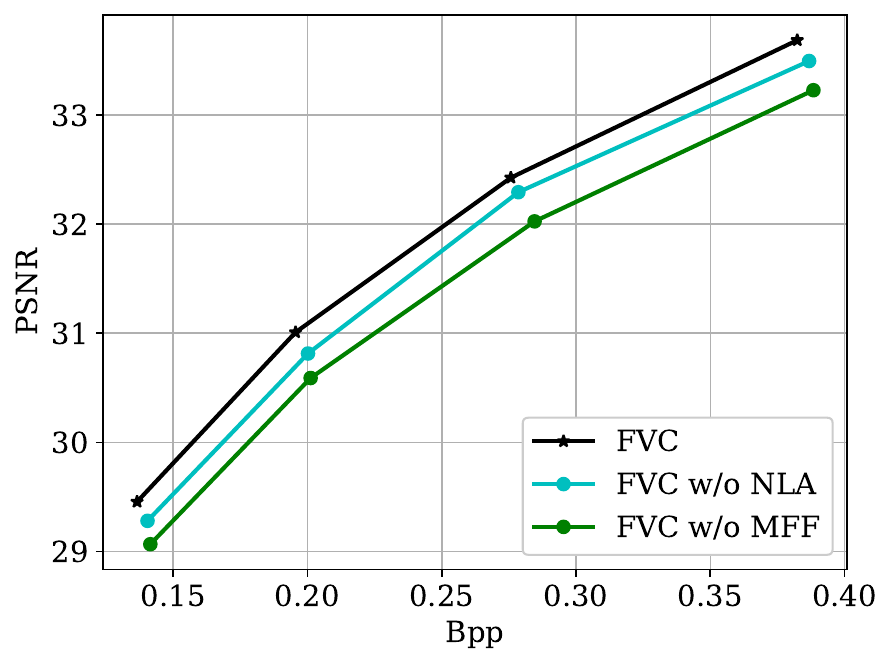}\label{fig:abmff}}
    % \vspace{-mm}
    \end{minipage}%
    \begin{minipage}[t]{0.5\linewidth}
    \centering
    \subfigure[]{\includegraphics[height=1.23in]{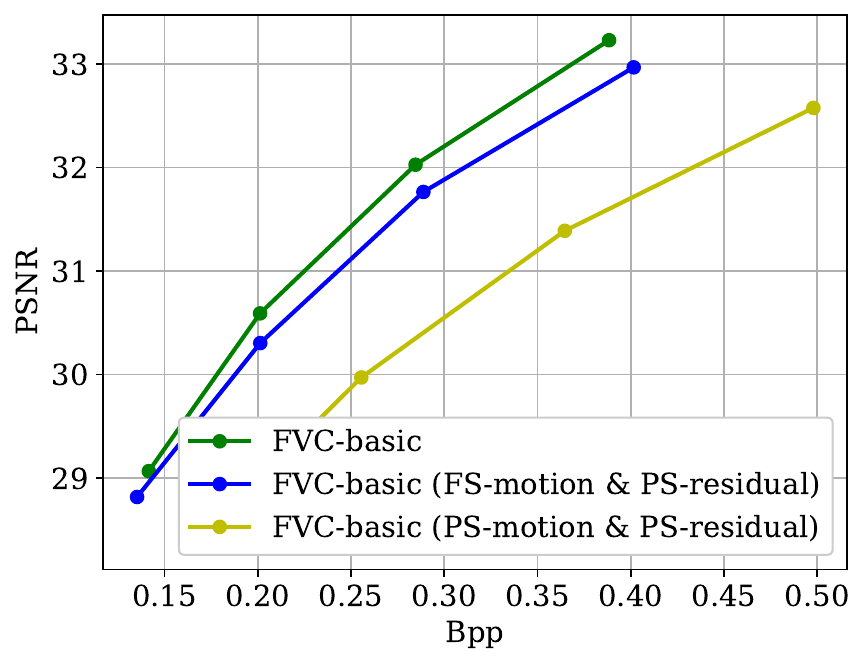}\label{fig:abfea}}
    % \vspace{-1mm}
    \end{minipage}
%   \includegraphics[height=1.2in]{figures/ablation/HEVC_Class_D_mff_psnr.pdf}%
%   \includegraphics[height=1.2in]{figures/ablation/HEVC_Class_D_fea_psnr.pdf}
% \end{center}
  \caption{Ablation study on the HEVC Class D dataset. (1) \textbf{FVC}: Our proposed method FVC. (2) \textbf{FVC w/o NLA}: FVC without adopting the non-local attention (NLA) mechanism in the multi-frame feature fusion (MFF) module. (3) \textbf{FVC w/o MFF (also referred to as FVC-basic)}: FVC without using the whole MFF module. (4) \textbf{FVC-basic (FS-motion \& PS-residual)} and (5) \textbf{FVC-basic (PS-motion \& PS-residual)} are two variants of FVC-basic by only performing residual compression at the pixel space and by performing both motion-related operations and residual compression at the pixel space, respectively. Note the motion-related operations and residual compression in FVC-basic are both performed in the feature space, which are referred to as FS-motion and FS-residual, respectively. We also refer to the corresponding pixel-space operations in DVC* as PS-motion and PS-residual.}
  
\label{fig:ablation}
\end{figure}

\textbf{Effectiveness of Different Components.} As shown in Fig.~\ref{fig:ablation},  we take the HEVC Class D dataset as an example to demonstrate the effectiveness of different modules in our proposed framework FVC. 
As shown in Fig.~\ref{fig:abmff}, to demonstrate the effectiveness of the non-local attention (NLA) module, we remove NLA in the multi-frame feature fusion (MFF) module and the results of FVC w/o NLA drop nearly 0.2dB when compared with our complete model (\textit{i.e.}, FVC).
We also introduce the basic model of our approach~(\textit{i.e.}, FVC w/o MFF or FVC-basic), where the multi-frame feature fusion module is removed in our complete model. When compared with our complete model (\textit{i.e.}, FVC), the result of our basic model FVC-basic after removing the MFF module drops 0.5dB at 0.3bpp~(see the green and black curves). These experimental results demonstrate the effectiveness of our newly proposed MFF module equipped with the NLA mechanism for fusing the features from multiple previous frames.

% To verify the effectiveness of our proposed feature-level operations, we replace the feature-level motion-related operations (\textit{i.e.}, the deformable compensation module) (FL-motion) and feature-level residual compression module (FL-residual) by the corresponding pixel-level motion-related operations (PL-motion) and pixel-level residual compression (PL-residual) in DVC*. Here, the motion compensation network is removed for fair comparisons.
% After replacing our FL-residual module with PL-residual, FVC-basic (FL-motion \& PL-residual) outperforms FVC-basic (PL-motion \& PL-residual) with all operations performed in pixel space and achieves 1.2dB improvement at 0.4bpp, which proves the effectiveness of our proposed deformable compensation module.

To verify the effectiveness of our proposed feature-level operations, in Fig.~\ref{fig:abfea} we report the results of two variants of our FVC-basic.
\textbf{(1) FVC-basic (FS-motion \& PS-residual)}. The predicted feature $\bar{F}_t$ after our deformable compensation module is converted as the predicted frame by using our frame reconstruction module, which is then used as the input to the pixel-space residual compression module in DVC*.
\textbf{(2) FVC-basic~(PS-motion \& PS-residual}. We perform both motion-related operations and residual compression at the pixel space, which is conceptually the same as DVC* except some subtle differences in implementation details.

It is noted that FVC-basic outperforms FVC-basic~(FS-motion \& PS-residual) by 0.3dB at 0.38bpp, which indicates it is beneficial to perform residual compression at the feature space. Furthermore, FVC-basic~(FS-motion \& PS-residual) achieves 1.2dB improvement at 0.4bpp when compared with FVC-basic~(PS-motion \& PS-residual), which demonstrates it is also necessary to perform motion compensation at the feature space.

\begin{figure}[t]
\centering
% \begin{center}
  \includegraphics[width=0.57\linewidth]{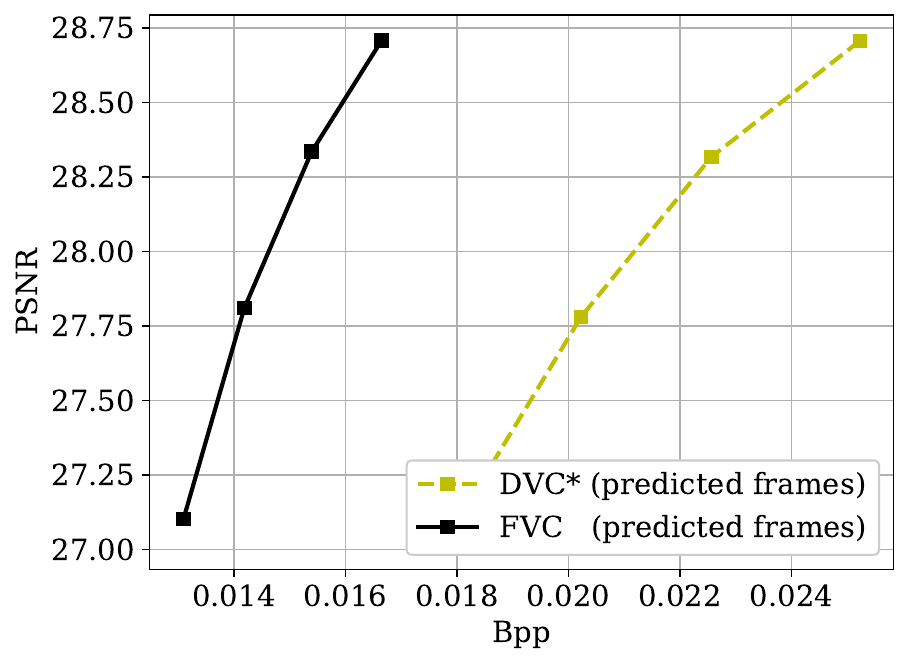}
% \vspace{-5mm}
% \end{center}
  \caption{Average PSNR(dB) over all predicted frames and bpps used to encode motion information of our FVC and DVC* on the HEVC Class C dataset.}
\label{fig:motionab}
\end{figure}

\textbf{Analysis of Deformable Compensation.}
For better comparison of the motion compensation module in the feature space and the pixel space, we also take the predicted feature $\bar{F}_t$ as the input of our frame reconstruction module to produce the predicted frame and evaluate the PSNR of the predicted frame and the corresponding bpp for compressing motion information. As shown in Fig.~\ref{fig:motionab}, the predicted frames of our proposed FVC achieves 1.75dB improvement at 0.017bpp on the HEVC Class C dataset when compared with the predicted frames of DVC*. It should be mentioned that DVC* uses the same compression method based on the same auto-encoder style network as our proposed method FVC, so the result clearly demonstrates the effectiveness of our proposed deformable compensation module.

\begin{figure}[!t]
\centering
\begin{minipage}[t]{0.5\linewidth}
\centering
\subfigure[Frame No.6 (Ground-truth)]{\includegraphics[width=0.9\linewidth]{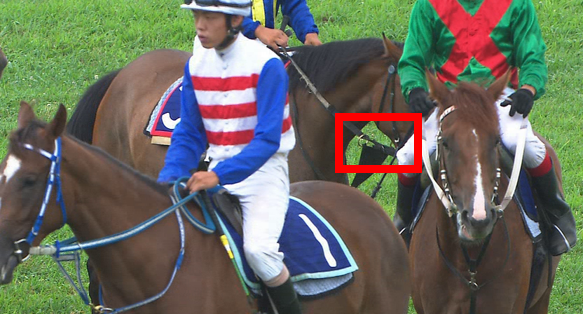}\label{fig:frame6}}
\end{minipage}%
\begin{minipage}[t]{0.5\linewidth}
\centering
\subfigure[Optical flow map (DVC*)]{\includegraphics[width=0.9\linewidth]{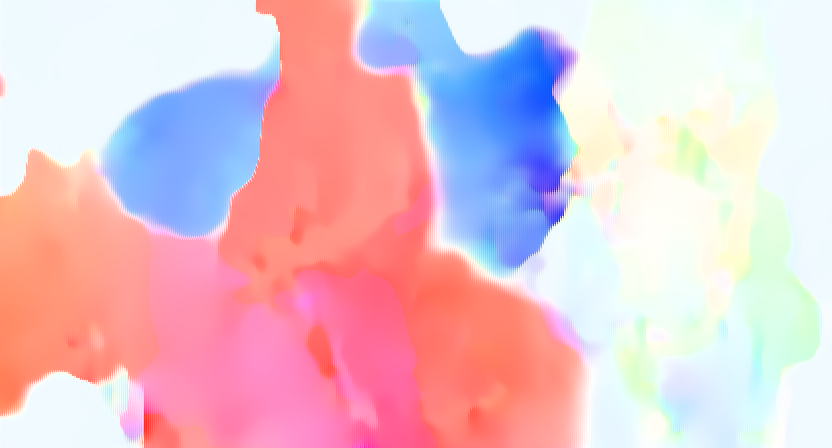}\label{fig:opticalflow}}
\end{minipage}
\begin{minipage}[t]{0.5\linewidth}
\centering
\subfigure[Offset map 1 of DC (Ours)]{\includegraphics[width=0.9\linewidth]{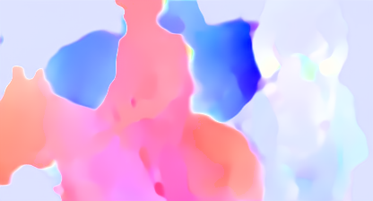}\label{fig:offset1}}
\end{minipage}%
\begin{minipage}[t]{0.5\linewidth}
\centering
\subfigure[Offset map 2 of DC (Ours)]{\includegraphics[width=0.9\linewidth]{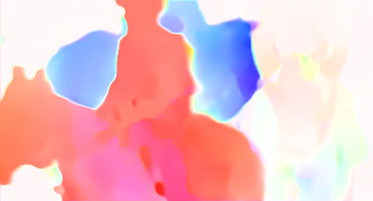}\label{fig:offset2}}
\end{minipage}
\begin{minipage}[t]{0.3333333333333\linewidth}
\centering
\subfigure[One patch (GT)]{\includegraphics[width=0.99\linewidth]{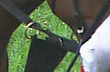}\label{fig:gtball}}
\end{minipage}%
\begin{minipage}[t]{0.3333333333333\linewidth}
\centering
\subfigure[Flow (26.68/0.0346)]{\includegraphics[width=0.99\linewidth]{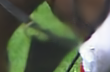}\label{fig:flowball}}
\end{minipage}%
\begin{minipage}[t]{0.333333333333\linewidth}
\centering
\subfigure[DC (26.95/0.0223)]{\includegraphics[width=0.99\linewidth]{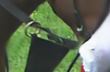}\label{fig:dcball}}
\end{minipage}
\caption{Visualization of the results by using optical flow based motion compensation (b,f) and our deformable compensation (c,d,g) for the 6th frame (a) in the \textit{RaceHorses} sequence from the HEVC Class C dataset. (b): the reconstructed optical flow map from the motion compression network of DVC*. (c,d): two representative reconstructed offset maps used for deformable convolution (DC) in our deformable compensation module. Visualization results of one zoomed-in patch from: the ground-truth (e), the flow based motion compensation module from DVC* (f) and our deformable compensation module (g), in which both PSNR and bpps for coding motion information are reported.}
\label{fig:offsetvis}
\end{figure}

\textbf{Visualization of Deformable Compensation Results.}
In Fig.~\ref{fig:offsetvis}, we take the 6th frame in the \textit{RaceHorses} sequence from the HEVC Class C dataset as an example and visualize motion information and the motion compensation results. We visualize the optical flow map used for motion compensation in DVC* (see Fig.~\ref{fig:opticalflow}) and two representative offset maps (from the total number of $72$ offset maps) used for deformable convolution in our work (see Fig.~\ref{fig:offset1} and Fig.~\ref{fig:offset2}). It can be observed that the learned offset maps in our work encode similar motion patterns as the optical flow map from DVC*.  We also observe that the motion compensation result for one patch after using our proposed deformable compensation (DC) module are visually more similar to the ground-truth patch when compared with that by using the optical flow based motion compensation method in DVC*. In practice, we can achieve 0.27dB gain by only using 65\% bpp for motion coding (see Fig.~\ref{fig:gtball}, Fig.~\ref{fig:flowball} and Fig.~\ref{fig:dcball}).

\textbf{Running Time and Model Complexity.}
The total number of parameters in our proposed framework is about 26M, in which the parameters from the compression network used for offset maps and residual feature maps take more than 24M.
We use the videos with the resolution of $1920\times1080$ to evaluate the inference time on the machine with a single 2080TI GPU (11GB Memory). The coding time of our proposed frameworks FVC and FVC-basic as well as DVC and DVC* are 548ms, 201ms, 460ms and 709ms, and the corresponding BDBRs on the HEVC Class D dataset by using H.265 as the anchor method are -18.39\%, -7.09\%, 14.08\% and 4.31\%. From the results, we observe that our FVC-basic outperforms DVC and DVC* in terms of both efficiency and effectiveness. In addition, our proposed framework FVC achieves the best BDBR result with 347ms used for the multi-frame feature fusion module, which can be accelerated by using a simpler fusion module in our future work.

\vspace{-1mm}
\section{Conclusion}
\vspace{-1mm}
In this work, we have proposed a new framework FVC for deep video compression in the feature space, which consists of the deformable compensation module, the feature-level residual compression module and the multi-frame feature fusion module. The deformable compensation module first predicts the offset maps as motion information, which are then compressed by using the newly proposed motion compression module and the reconstructed offset maps will be finally used in deformable convolution for motion compensation. The proposed multi-frame feature fusion module takes multiple feature representations from the current frame and the previous frames and uses the deformable compensation and the non-local attention mechanism to refine the initial reconstructed feature for better frame reconstruction. By performing all the operations in the feature space, our framework achieves promising results on the HEVC, UVG, VTL and MCL-JCV datasets.% Based on our proposed framework, more operations like bi-directional prediction, multi-scale motion compensation can be performed to further improve the performance in our future work.

\noindent\textbf{Acknowledgement}
This work was supported by the National Key Research and Development Project of China (No. 2018AAA0101900). 

{\small
\bibliographystyle{ieee_fullname}
\bibliography{egbib}
}

\end{document}